\documentclass[journal]{IEEEtran}
\usepackage{graphicx}
\usepackage{booktabs}
\usepackage{tabularx}
\usepackage{cite}
\usepackage{url}
\usepackage[T1]{fontenc}
\usepackage[utf8]{inputenc}

\begin{document}
\title{A Token Service Interface for AI-Native RANs}
\author{Jianan Zhang,~\IEEEmembership{Member,~IEEE}, Peng Li,~\IEEEmembership{Graduate Student Member,~IEEE}, and Xiang Cheng,~\IEEEmembership{Fellow,~IEEE}
\thanks{The authors are with the School of Electronics, Peking University, China.}}
\maketitle

\begin{abstract}

Generative and embodied AI services exchange token streams within continuing inference and control loops. Their communication requirements depend on each token set's purpose, useful timing, and execution context. This article organizes these properties into service, temporal, and stateful semantics and proposes a token service interface (TSI) between applications, radio access networks (RANs), and edge runtimes. TSI binds delivery requirements, readiness forecasts, and execution-state references to each schedulable token set, specifying field ownership, versioned updates, and admission feedback. A drone inspection case study over a decoupled RAN illustrates importance-aware radio allocation, advance preparation for timely delivery, and selective state migration that balances interruption against forwarding delay.

\end{abstract}


\section{Introduction}
\label{sec:introduction}

Artificial intelligence (AI) is shifting wireless traffic toward representations produced and consumed by models. A vision-language-action (VLA) service exchanges observations, language context, and control outputs across successive rounds \cite{kim2024openvla}. Model computation can alternate between burst and trickle phases, assign unequal values to transmitted units, and retain state whose loss wastes compute. Within one task, observations can tolerate selective delivery, an action command can require complete delivery, and execution state can remain valuable across radio handovers. These requirements can differ even when average rate and delay targets match. In this article, an \emph{AI service} exchanges tokenized representations while maintaining an inference or control objective across communication rounds. An AI-native radio access network (RAN) accounts for such service requirements in communication decisions, complementing the use of AI within RAN functions. 

Conventional flow descriptors provide little context about whether partial delivery is useful, future emissions are predictable, or successive sets depend on retained execution state. Task-oriented and token communication optimize representations and importance \cite{liu2024adaptable,lei2026importance,qiao2025tokcom}. TokCom also discusses exposing task semantics and retaining inference state under mobility \cite{fu2026tokcom}. Release~18 protocol data unit (PDU) sets support importance, delay budgets, error rates, and integrated handling indicating whether all constituent PDUs are needed \cite{ts23501r18}. Decoupled access addresses uplink-downlink asymmetry \cite{boccardi2016decouple}, while serving systems manage inference phases and retained state \cite{zhong2024distserve,kwon2023vllm}. This article develops a token-set descriptor that binds application delivery requirements, runtime readiness forecasts, and execution-state references, with explicit field ownership and update rules. 
The feasibility of a delivery target depends on both its timing requirement and the path to the execution context. These constraints should therefore be interpreted together.

We organize the communication-relevant properties of an AI service into three dimensions: \emph{service semantics} state what outcome is requested, \emph{temporal semantics} state when it matters, and \emph{stateful semantics} state which execution context it belongs to. Token-set granularity allows delivery requirements to vary within one session. The application requests an outcome; the radio access network (RAN) and edge runtime translate the request into protection, scheduling, and placement decisions and return the service that they can support.

The article first develops the three semantic dimensions and distinguishes measured task outcomes, application-aware delivery surrogates, and network performance metrics. It then specifies the descriptor for the semantics, field ownership, and interface operation. Finally, proof-of-concept comparisons illustrate how this information affects network decisions using a drone inspection setting over a decoupled RAN that separates uplink and downlink base stations. Decoupling improves the uplink, while importance, advance timing, and retained state inform allocation, scheduling, and placement choices.

\section{Communication Semantics of AI Services}
\label{sec:semantics}

Communication semantics are application properties that change communication decisions. A \emph{token} is an application representation unit, such as a visual patch, language symbol, or action component; a \emph{token set} groups units sharing a delivery context. One set may span several packets. Selective treatment requires sets to remain identifiable and independently schedulable. All three dimensions apply across perception, reasoning-related exchanges, and actions.

\subsection{Service Semantics}

Service semantics state \emph{what} communication service is requested. Perception can produce data-heavy bursts from a power-limited device. Language and reasoning exchanges can be sequential and tightly ordered. Action outputs can be small yet critical. One VLA session can carry all these roles, so traffic class alone cannot describe its delivery requirements~\cite{kim2024openvla}.

Token sets contribute unequally: an inspected component matters more than background, although importance changes with the task and scene. A \emph{soft} regime permits selective transmission when partial evidence remains useful. A \emph{hard} set, such as a structured action command, requires complete, correctly ordered reconstruction. These are delivery requirements rather than modality labels: perception may require complete delivery, and action streams may contain optional updates.

For an additive utility profile, expected delivered importance is obtained by multiplying each token set's weight by its delivery probability and summing over the sets in a specified collection. Dividing by the total weight of that collection gives normalized expected utility. Delivery after an applicable deadline does not count.

Evaluation follows three levels: measured task outcomes; application-aware delivery surrogates, such as importance-weighted timely delivery; and network metrics, such as reliability, latency, and cost. Saliency can provide token-set importance weights, but the resulting delivery metric remains a surrogate unless it is calibrated and validated against task outcomes. Separating these levels allows recalibration without redesigning the controller.

\subsection{Temporal Semantics}

Temporal semantics state \emph{when} the service matters. Autoregressive inference first processes input context during prefill and then generates output tokens sequentially during decode. Time to first token measures initial responsiveness; time per output token measures subsequent cadence \cite{zhong2024distserve}. Input bursts followed by output trickles create directional demands that average rates miss. A deadline identifies the latest useful delivery time, while useful lifetime captures value decay. 

A system runtime is execution software on a device or at the edge that produces token sets or manages retained state. Schedulers can learn arrival patterns and prepare grants in advance \cite{elkael2026auguste}. The system runtime can additionally forecast aperiodic emissions from execution progress, with validity intervals and confidence. Such forecasts can reveal information that traffic observations miss before emission and enable preparation for prompt transmission upon arrival. Timely delivery of selected perception sets can be more useful than complete delivery after a control deadline.

\subsection{Stateful Semantics}

Stateful semantics state \emph{which} execution context the token set belongs to. A session retains earlier computation: the key-value (KV) cache stores attention states for reuse during generation \cite{kwon2023vllm}, while embodied agents retain history across perception-action rounds. Losing access to the serving model instance may require state transfer or repeated prefill, stalling inference even when the interrupted set is small. Session identity and compute affinity identify where successive sets should continue.

\begin{figure*}[t]
\centering
\includegraphics[width=\textwidth]{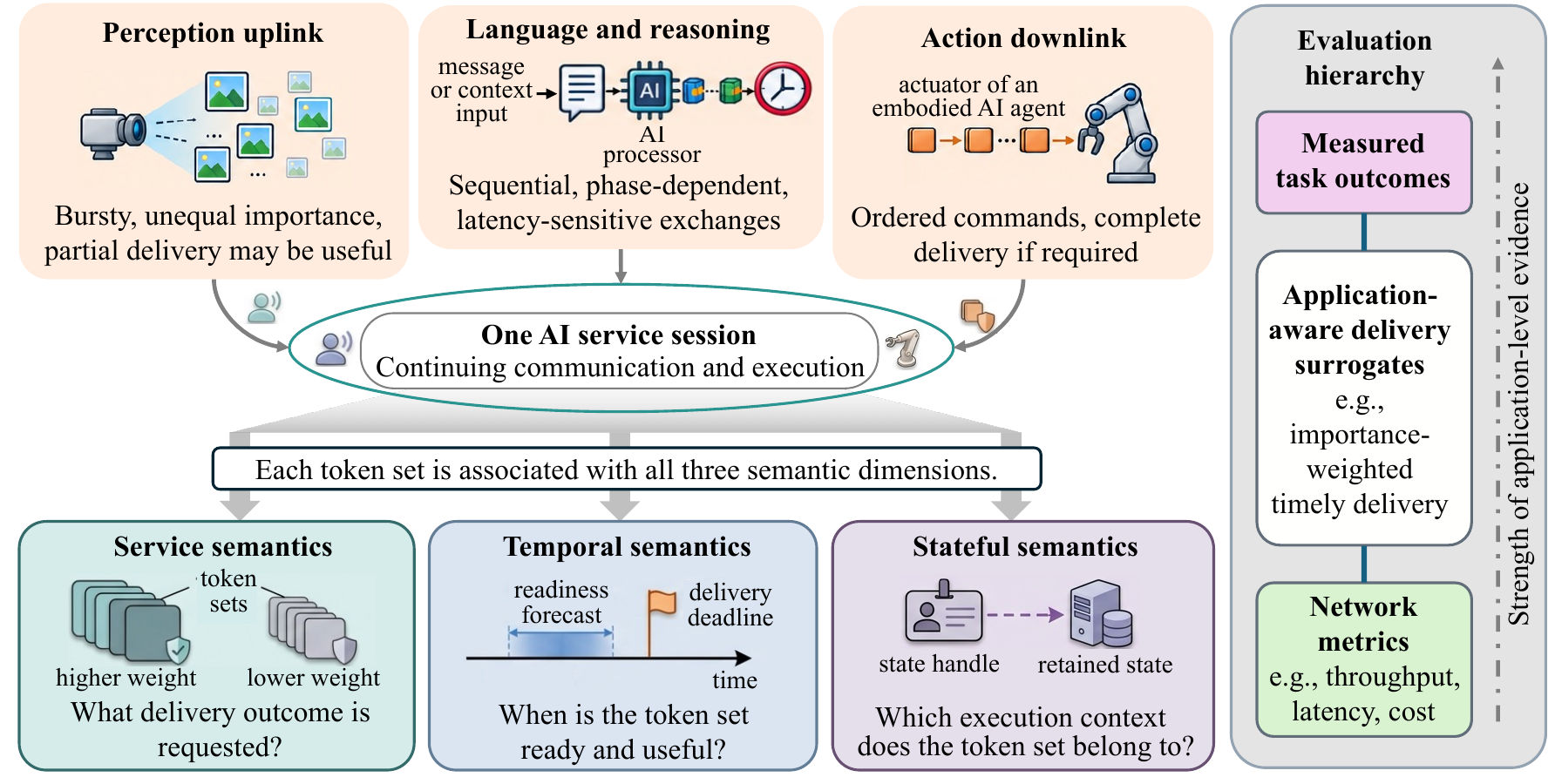}
\caption{Communication semantics and evaluation hierarchy for one AI service session.}
\label{fig:semantics}
\end{figure*}

Radio association and execution-state placement can evolve separately. After handover, requests may be forwarded to the current compute instance, or state may migrate when subsequent forwarding savings justify the migration cost. Figure~\ref{fig:semantics} summarizes the three dimensions and evaluation hierarchy.

\section{Token Service Interface}
\label{sec:tsi}

\begin{figure*}[t]
\centering
\includegraphics[width=\textwidth]{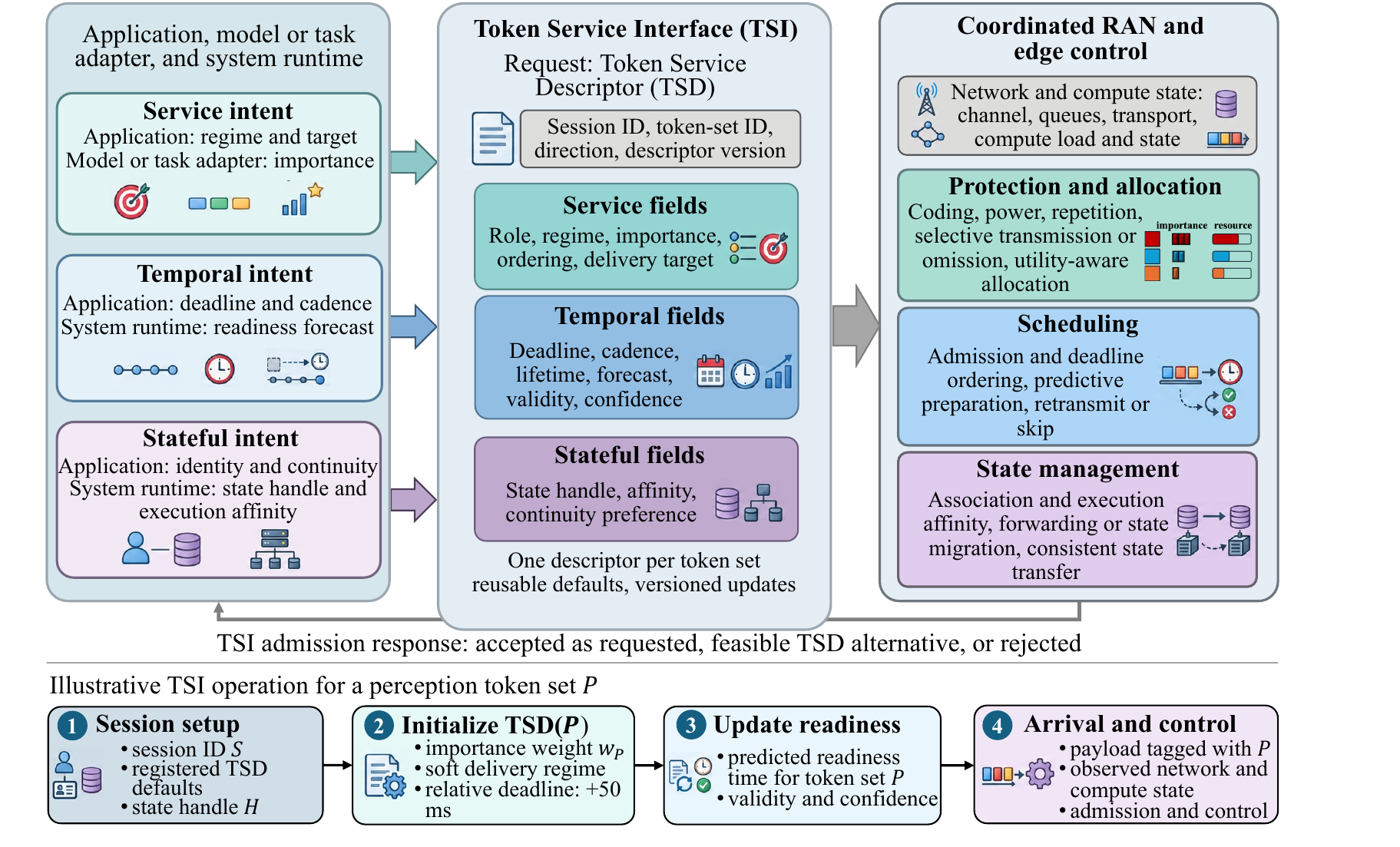}
\caption{Bidirectional TSI operation. Field owners initialize and update TSD. Coordinated RAN and edge controllers jointly interpret the TSD and return an admission result. The lower sequence shows an operation example.}
\label{fig:tsi}
\end{figure*}

The \textbf{Token Service Interface (TSI)} makes these semantics available to the RAN and the edge runtime. It serves as a boundary between application and network, through which the field owners export a \textbf{Token Service Descriptor (TSD)} for each schedulable token set, and the controllers return an admission result. Figure~\ref{fig:tsi} illustrates how the boundary turns the three semantic dimensions into coordinated decisions.

\subsection{Architectural Principles and Existing Mechanisms}

TSI exports communication-relevant properties while keeping model internals and task content private. This separation allows models, tokenizers, and serving pipelines to evolve independently of the interface. 

The application, model or task adapter, and system runtime supply their respective fields. The RAN controls radio admission and resource allocation, while edge control manages execution placement and consistent state transfer. Using observed network and compute state, they accept the requested TSD, propose a feasible alternative, or reject it. The application can reduce load, relax targets, adjust token-set granularity, or postpone generation.

TSI extends the information available through existing
mechanisms. Release~18 defines per-set treatment through PDU-set quality-of-service (QoS) parameters in a flow profile \cite{ts23501r18}. TSI builds on this per-set approach with AI-native fields in three respects. First, requested deadline and reliability targets can be declared per token set, whereas Release~18 carries at most one PDU-set delay budget and error rate per QoS flow direction. Second, system runtimes supply execution-aware readiness forecasts. Existing time-sensitive communication assistance information (TSCAI) provides flow-level periodicity and burst-arrival assistance \cite{ts23501r18}. TSI binds runtime-generated forecasts, their validity, confidence, and updates to individual sets for aperiodic transmissions. Third, the stateful dimension exposes execution affinity and continuity requirements for coordinated association, forwarding, and migration. Building on TokCom's architectural vision \cite{fu2026tokcom}, TSI specifies the common token-set binding, field ownership, update rules, and admission response that RAN and edge controllers use together.

\subsection{Token Service Descriptor and Field Ownership}

A TSD identifies the session, token set, direction, payload extent, and descriptor version. The token-set identifier binds metadata to packetized data. Reusable defaults registered at session setup cover slowly changing fields; each per-set descriptor references those defaults and supplies its overrides.

\textbf{Service fields} include role, soft or hard delivery regime, importance, ordering, and requested utility or completion reliability. The model or task adapter owns the importance field because it estimates the token set's contribution to the application objective. The application owns the regime and target fields because it can declare whether partial output remains useful. The TSD carries the token set's importance as a weight or class. If an implementation registers a reusable importance table, the TSD may instead reference the table version and entry. Aggregate utility targets apply to specified collections of token sets.

\textbf{Temporal fields} include deadline, cadence, useful lifetime, and optional readiness forecasts with validity, confidence, and lead time. The application owns deadline and cadence fields. The system runtime producing the token set owns forecasts because it observes execution progress and can update its forecast before emission. A reference event and clock domain determine each set's absolute deadline and residual budget. Deadline policies may persist across emissions, whereas forecasts are revisable estimates for advance preparation.  

\textbf{Stateful fields} include session identity, an opaque state handle, execution affinity, and continuity preference. The application supplies identity and continuity requirements. The system runtime managing the retained state supplies the state handle and execution affinity, and resolves the handle to the state's placement, size, and availability. Network measurements supply transport delay and transfer bandwidth. Owners update these fields when requirements or state change; the RAN need not inspect model state.

A token set may contain one or multiple tokens. Finer granularity enables selective allocation but increases signaling. Quantized importance classes and token aggregation balance selective allocation against signaling overhead. Token-set identifiers associate packets with their set. Token-set decisions must be mapped to radio transport blocks during implementation.

\subsection{From Semantics to Decisions}

The three dimensions primarily inform three decision classes. \textbf{Service semantics primarily inform protection and allocation}: coding, power, repetition, selective transmission, and utility-aware allocation. Hard action sets may receive uniform strong protection, while allocation for soft perception traffic favors high-value token sets. \textbf{Temporal fields primarily inform scheduling}: admission, deadline ordering, predictive reservation, and timely retransmission. \textbf{Stateful fields primarily inform state management}: association, forwarding to the serving model instance, compute affinity, and migration under mobility. 

Despite the primary associations, these decisions remain coupled. Stronger protection takes additional time, and migration can reduce the budget for an arriving token set. Under a fixed radio budget, a robustness view maximizes normalized expected utility. Under a fixed utility target, an efficiency view minimizes radio resources. 

We next describe TSI operation. At session setup, the service establishes session ID $S$, registers TSD defaults, and associates state handle $H$. For a forthcoming token set $P$, the field owners initialize TSD($P$) with its token-set identifier, direction, service fields, and deadline. The system runtime may then update TSD($P$) with a readiness forecast, validity interval, and confidence. When packets arrive tagged with $P$, the controllers combine the latest TSD($P$) with observed network and compute state for admission and coordinated communication and computation resource control.

Role and regime may persist across an inference phase, while importance can change with each scene or emission. Temporal fields typically change per emission or control cycle, and stateful fields change per session or compute-placement event. Owners issue versioned updates when application requirements, placement, retained state, or forecasts change. Newer metadata supersedes earlier versions, expired forecasts are ignored, and missing optional fields invoke agreed defaults.


\section{Illustrative Evaluation in a Decoupled RAN}
\label{sec:realization}

\subsection{Architecture and Service Profiles}
\label{sec:heterogeneity}

\begin{figure*}[t]
\centering
\includegraphics[width=\textwidth,trim=0 0 0 0,clip]{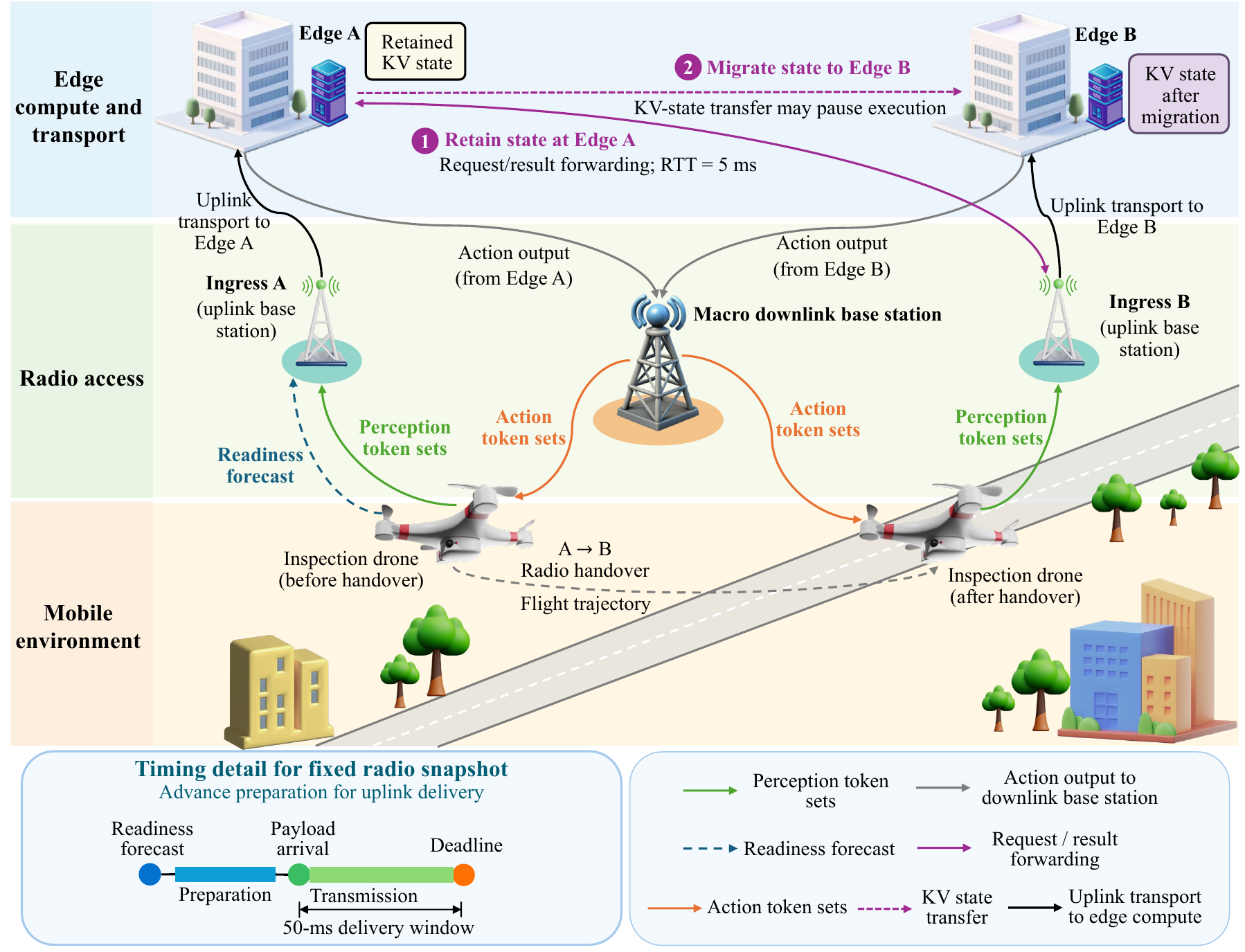}
\caption{Inspection drone operation with decoupled uplink, downlink, and edge execution. After radio handover from Ingress $A$ to Ingress $B$, execution state can remain at Edge $A$ with request and result forwarding or migrate to Edge $B$.}
\label{fig:decoupled}
\end{figure*}

The drone uses a nearby uplink base station as perception receiver and a macro downlink base station to transmit actions under decoupled access \cite{boccardi2016decouple}, as illustrated in Fig.~\ref{fig:decoupled}. We evaluate the dimensions separately. Service and temporal semantics use a fixed radio snapshot, and stateful semantics use a prescribed flight trajectory. Joint control is outside this illustration's scope. In this snapshot, the macro and nearby uplink base stations are 250~m and 60~m from the drone, respectively. An assumed 23-dBm transmit power, $-95$-dBm noise power, and path loss $35+35\log_{10}(d/1\,\mathrm{m})$~dB give the SNRs in Table \ref{tab:model}. The service evaluation compares equally weighted and importance-aware profiles under coupled and decoupled access. The temporal evaluation assesses the utility benefit of advance preparation. The stateful evaluation compares forwarding and migration policies using a common drone trajectory, handover sequence, and request arrivals.

\begin{figure*}[t]
\centering
\includegraphics[width=\textwidth]{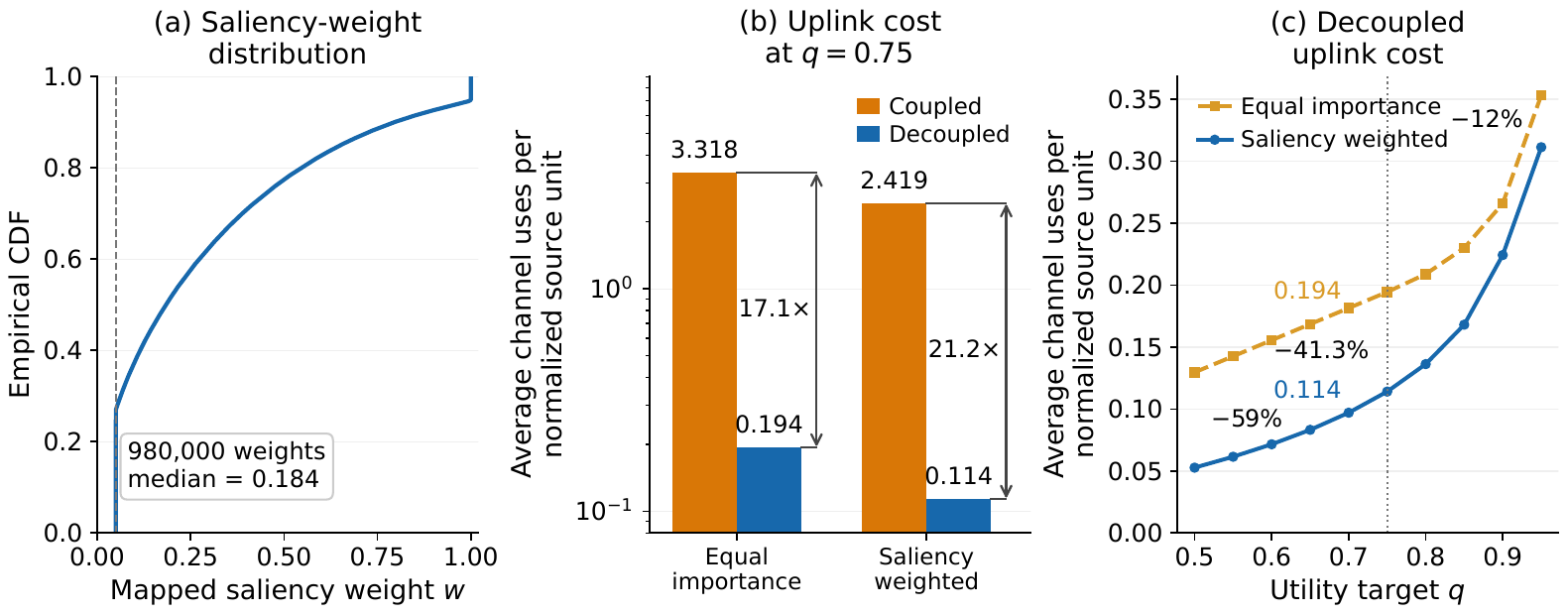}
\caption{Service evaluation under distinct utility profiles: (a) COCO saliency-weight distribution; (b) optimized uplink cost for equal-importance and saliency-weighted profiles; (c) decoupled uplink cost versus utility target, with relative cost reductions between profiles. Cost is averaged over all original
source units with each unit’s payload normalized to one bit.}
\label{fig:evidence}
\end{figure*}

We use the 5,000 COCO 2017 validation images \cite{lin2014coco} to derive an importance profile. Each image is resized to $224\times224$ and divided into 196 patches. Spectral-residual saliency \cite{hou2007saliency} is averaged per patch, clipped at the 5th and 95th percentiles, and mapped to $[0.05,1]$. The top 20\% of weights carry 53.4\% of importance (Fig.~\ref{fig:evidence}(a)).

\subsection{Resource Savings with Service Semantics}
\label{sec:matched-validation}

Equal importance and saliency weighting share Table~\ref{tab:model}'s settings. The saliency-weighted profile is a representative synthetic vector of 196 midpoint quantiles of the pooled weights. Each source unit is a singleton token set. The target $q$ specifies the minimum normalized expected utility, defined as the expected sum of the weights of successfully delivered sets divided by the total weight of all original sets. Thus, $q=0.75$ requires delivery of at least 75\% of the total importance on average. Each profile normalizes utility using its own weights. An outage model maps channel uses to delivery probabilities under independent Rayleigh blocks at the tabulated average SNR. A solver deterministically selects token sets and allocates channel uses to minimize cost while meeting the expected weighted target. Cost is averaged over all original source units, with each unit's payload normalized to one. The model excludes finite-blocklength effects, retransmissions, and signaling overhead. Only uplink resource cost is compared, and the common macro downlink is fixed.

\begin{table}[t]
\centering
\caption{Evaluation settings.}
\label{tab:model}
\scriptsize
\renewcommand{\arraystretch}{1.13}
\begin{tabularx}{\columnwidth}{p{0.26\columnwidth} X}
\toprule
\textbf{Item} & \textbf{Setting}\\
\midrule
Token transmission & 196 equal-size singleton token sets; target $q=0.75$ (swept in Figs. \ref{fig:evidence}(c) and \ref{fig:temporal-stateful}(a)).\\
Radio snapshot & Average uplink SNR: $-0.93$~dB (coupled), 20.76~dB (decoupled); no interference.\\
Temporal window & 50~ms; preparation fractions $\eta=$ 10\%, 30\%, and 50\%.\\
Stateful requests & 1,228 arrivals over 122.765 s at the ingress; 100-ms period; 30-ms inference.\\
Drone mission & Hover 10~s, two 40-m-radius orbits, relocation, hover 30~s; altitude 40~m throughout; speed cap 10~m/s.\\
Sites & Two ingress sites A and B, 200~m apart and 10~m high, each with its own edge execution site; initial ingress and execution anchor at A.\\
Handover & Equal-power line-of-sight scores; 3-dB advantage sustained for 160~ms. Uplink ingress sequence $A\to B\to A\to B\to A\to B$; physical handovers at 24.83, 35.16, 49.96, 60.29, and 82.30~s.\\
Forwarding & Additional transport RTT 5~ms.\\
Retained state & 56~KiB/token; reference 8,192 tokens (448~MiB); plotted range 2k--16k.\\
Migration & 10~Gb/s KV state transfer between compute instances; model weights preloaded.\\
\bottomrule
\end{tabularx}
\end{table}

We first examine how heterogeneous token importance affects the uplink resource savings achieved by decoupling. The higher uplink SNR in the decoupled RAN reduces the channel uses needed to meet a given delivery target. At $q=0.75$, the saliency-weighted profile requires 27.1\% less uplink resource than the equal-importance profile under coupled access and 41.3\% less under decoupled access. The coupled-to-decoupled cost ratios are 17.1 and 21.2, respectively (Fig.~\ref{fig:evidence}(b)), making the ratio 24.3\% higher with saliency weighting. The stronger uplink supports more reliable delivery of selected token sets: the weighted allocation protects 93 high-value token sets under decoupled access and 133 under coupled access, meeting the same weighted target with fewer active token sets.

Figure~\ref{fig:evidence}(c) compares the profiles as the utility target varies. Under decoupled access, the relative cost difference between profiles decreases from 59\% at $q=0.5$ to 12\% at $q=0.95$. Higher targets leave less scope to omit low-value token sets, so resource savings depend on both the utility target and radio conditions. 

\subsection{Performance Improvements with Temporal and Stateful Semantics}
\label{sec:session}

\textbf{Temporal semantics.} We hold the saliency-weighted profile fixed and compare achievable utility with and without advance preparation. The minimum per-cycle budget $C^*(q)$ for utility $q$ is provisioned over a 50-ms window from payload arrival to the delivery deadline. An exact readiness forecast with sufficient lead time allows grant and resource preparation to finish before packet arrival and remain valid for transmission. Without this forecast, preparation consumes a fraction $\eta$ of the window, leaving $(1-\eta)C^*(q)$ usable at the same provisioned rate and deadline.

\begin{figure*}[t]
\centering
\includegraphics[width=\textwidth]{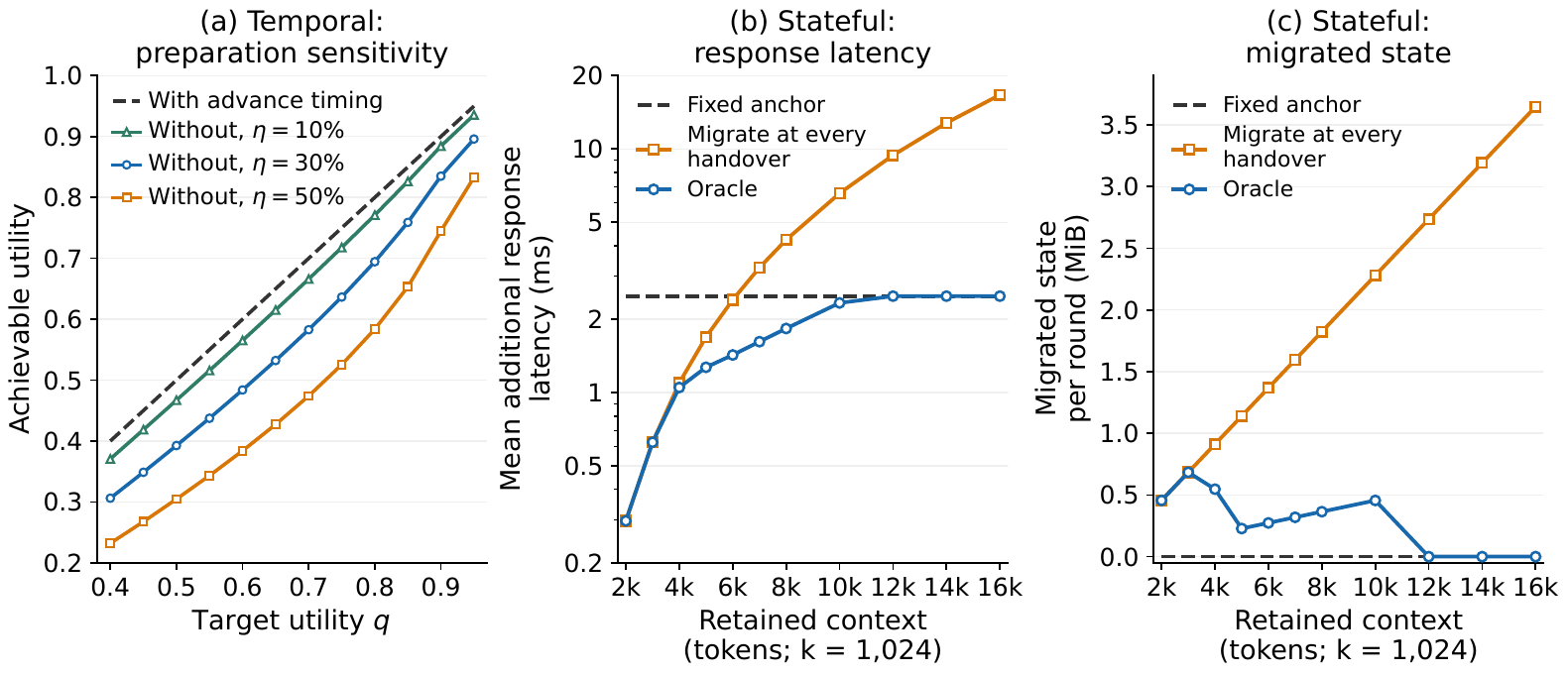}
\caption{(a) Temporal sensitivity to preparation duration under equal provisioned rate and deadline. Stateful evaluation: (b) mean additional response latency, on a logarithmic scale; (c) migrated state per completed round. Context spans 2k-16k linearly, with k = 1,024. Non-monotone oracle traffic reflects
changes in migration count; no migrations are selected at the plotted contexts of 12k and above.}
\label{fig:temporal-stateful}
\end{figure*}

With advance timing, the dashed line in Fig.~\ref{fig:temporal-stateful}(a) attains each target. Without it, utility falls as preparation consumes transmission time, despite reoptimizing allocation. Increasing the preparation fraction from 10\% to 50\% widens the gap. At $q=0.75$ and $\eta=0.3$, utility is 0.6364, a 15.1\% relative loss.


\textbf{Stateful semantics.} We compare fixed anchoring, migration at every handover, and an offline oracle on the common drone trace in Table~\ref{tab:model}. The oracle knows future requests and handovers and minimizes response latency over handover-associated retain or migrate decisions, breaking ties by fewer migrated bytes. It provides an ideal benchmark within this decision model. We assume a fixed 30 ms computation time per request, with arrivals every 100 ms and first-in, first-out (FIFO) processing.

Remote execution adds an assumed 5-ms forwarding round-trip delay, split equally before and after computation, representing transport and node processing. Its magnitude is consistent with the edge transport RTT range discussed in \cite{alriksson2020critical}. Migration waits for earlier requests to finish computing, then pauses execution for state transfer. Arriving requests wait and subsequently undergo FIFO processing. Additional latency measures ingress-to-result time minus computation, comprising queueing and extra forwarding but excluding radio transmission. We size the retained context of a history-preserving edge reasoning component using Qwen2.5-7B: 28 layers, 4 KV heads per layer, and 128 elements per head. Storing keys and values at 16 bits gives 56 KiB per token \cite{qwen2024config}. Retained context size is fixed within each run. At 8,192 context tokens, 448~MiB of KV state takes 378.3~ms to migrate at 10~Gb/s, including 2.5~ms of one-way transport. Model weights are preloaded, and additional cutover overhead is excluded.

At this context size, mean additional latencies are 2.488, 4.238, and 1.832~ms for fixed anchoring, every-handover migration, and the oracle, respectively (Fig.~\ref{fig:temporal-stateful}(b)). The oracle forwards during the two brief visits and migrates for the final residence, reducing additional latency by 26.4\% and 56.8\% relative to the baselines. Including computation, its edge-response latency reduction over every-handover migration is 7.0\%. Beyond this reduction in mean latency, the state-aware oracle improves service continuity by reducing migration-induced execution pauses from five to one, each lasting over 300 ms. Total migration traffic is 448~MiB versus 2,240~MiB (Fig.~\ref{fig:temporal-stateful}(c)). All policies preserve state and complete every request. Selective placement balances forwarding delay against migration stalls according to context size, forwarding delay, and request timing.

\section{Open Challenges}
\label{sec:challenges}

\textbf{Calibrating service semantics.} Saliency and linguistic roles provide proxies. Deleting or perturbing token sets and rerunning the downstream model can assess their contribution and calibrate delivery targets against task outcomes. Descriptors should convey calibration uncertainty; degradation under changing models or environments should trigger recalibration.

\textbf{Scaling temporal semantics.} Metadata, channel acquisition, clock alignment, scheduling, and retransmission costs determine useful token-set granularity and prediction horizons. The benefit of finer descriptors and advance preparation must justify signaling and processing overhead.

\textbf{Maintaining state under mobility.} Radio handover can outpace migration, and different controllers manage the two. Existing edge-serving policy transfers or partially recomputes KV state at handover \cite{lee2026kvhandover}. TSI specifies the fields and owners that such policies consume. Coordinated decisions require state size, transport conditions, compute queues, residence predictions, and consistent runtime ownership to avoid costly reactions to brief radio changes.  

\textbf{Deploying a shared interface.} Admission limits, authentication, and access control must protect resources, metadata, and state handles against inflated importance or urgency. Multi-agent deployments add interference and congestion. Privacy requires exposing useful service classes without revealing task content.

\textbf{Standardizing the boundary.} Building on Release~18 per-set QoS handling \cite{ts23501r18}, AI interoperability requires common meanings and ownership for delivery, timing, and continuity fields, with radio and edge protocol bindings that permit control mechanisms to evolve.

\begin{samepage}
\section{Conclusion}
\label{sec:conclusion}

This article proposes TSI as a bidirectional boundary through which field owners initialize and update a per-token-set TSD, while coordinated RAN and edge controllers return an admission result. Each TSD associates its token set with delivery requirements, temporal information, and execution context. These three semantic dimensions guide protection, scheduling, and state placement while preserving network authority over resource decisions. The case study illustrates how importance weighting changes the resource benefit of uplink–downlink decoupling, advance preparation preserves usable transmission time, and selective migration balances forwarding delay against migration interruption. The evaluations quantify network utility and cost under explicit assumptions. Establishing application-level benefits requires calibrating application-aware delivery surrogates against measured task outcomes and evaluating metadata overhead, prediction uncertainty, and coordination between radio and edge control.
\end{samepage}

\bibliographystyle{IEEEtran}
\bibliography{references09}

@article{liu2024adaptable,
  author  = {Liu, Chuanhong and Guo, Caili and Yang, Yang and Jiang, Nan},
  title   = {Adaptable Semantic Compression and Resource Allocation for Task-Oriented Communications},
  journal = {IEEE Trans. Cogn. Commun. Netw.},
  volume  = {10},
  number  = {3},
  pages   = {769--782},
  year    = {2024},
  doi     = {10.1109/TCCN.2023.3346481}
}

@inproceedings{zhong2024distserve,
  author    = {Zhong, Yinmin and Liu, Shengyu and Chen, Junda and Hu, Jianbo and Zhu, Yibo and Liu, Xuanzhe and Jin, Xin and Zhang, Hao},
  title     = {{DistServe}: Disaggregating Prefill and Decoding for Goodput-Optimized Large Language Model Serving},
  booktitle = {Proc. 18th USENIX Symp. Operating Systems Design and Implementation (OSDI)},
  pages     = {193--210},
  year    = {2024}
}

@inproceedings{kwon2023vllm,
  author    = {Kwon, Woosuk and Li, Zhuohan and Zhuang, Siyuan and Sheng, Ying and Zheng, Lianmin and Yu, Cody Hao and Gonzalez, Joseph E. and Zhang, Hao and Stoica, Ion},
  title     = {Efficient Memory Management for Large Language Model Serving with {PagedAttention}},
  booktitle = {Proc. 29th ACM Symp. Operating Systems Principles (SOSP)},
  pages     = {611--626},
  year      = {2023},
  doi       = {10.1145/3600006.3613165}
}

@article{qiao2025tokcom,
  author  = {Qiao, Li and Mashhadi, Mahdi Boloursaz and Gao, Zhen and Tafazolli, Rahim and Bennis, Mehdi and Niyato, Dusit},
  title   = {Token Communications: A Large Model-Driven Framework for Cross-Modal Context-Aware Semantic Communications},
  journal = {IEEE Wireless Commun.},
  volume  = {32},
  number  = {5},
  pages   = {80--88},
  year    = {2025},
  doi     = {10.1109/MWC.001.2500084}
}

@article{fu2026tokcom,
  author  = {Fu, Yaru and Ji, Liang and Maharjan, Sabita and Quek, Tony Q. S.},
  title   = {Token Communications ({TokCom}): A Unified {AI}-Native Communication Framework},
  journal = {arXiv preprint arXiv:2607.17628},
  year    = {2026},
  doi     = {10.48550/arXiv.2607.17628}
}

@article{lei2026importance,
  author  = {Lei, Kaiyi and Peng, Yuanzhe and Zhang, Letian and Xu, Jie},
  title   = {Importance-Aware Resource Allocation for Collaborative Task-Oriented Semantic Communication},
  journal = {arXiv preprint arXiv:2606.29052},
  year    = {2026},
  doi     = {10.48550/arXiv.2606.29052}
}

@article{boccardi2016decouple,
  author  = {Boccardi, Federico and Andrews, Jeffrey G. and Elshaer, Hisham and Dohler, Mischa and Parkvall, Stefan and Popovski, Petar and Singh, Sarabjot},
  title   = {Why to decouple the uplink and downlink in cellular networks and how to do it},
  journal = {IEEE Commun. Mag.},
  volume  = {54},
  number  = {3},
  pages   = {110--117},
  year    = {2016},
  doi     = {10.1109/MCOM.2016.7432156}
}

@inproceedings{kim2024openvla,
  author    = {Kim, Moo Jin and Pertsch, Karl and Karamcheti, Siddharth and others},
  title     = {{OpenVLA}: An open-source vision-language-action model},
  booktitle = {Proc. Conf. Robot Learning (CoRL)},
  series    = {Proc. Mach. Learn. Res.},
  volume    = {270},
  pages     = {2679--2713},
  year      = {2025}
}

@inproceedings{hou2007saliency,
  author    = {Hou, Xiaodi and Zhang, Liqing},
  title     = {Saliency detection: A spectral residual approach},
  booktitle = {Proc. IEEE Conf. Computer Vision and Pattern Recognition (CVPR)},
  pages     = {1--8},
  year      = {2007},
  doi       = {10.1109/CVPR.2007.383267}
}

@inproceedings{lin2014coco,
  author    = {Lin, Tsung-Yi and Maire, Michael and Belongie, Serge and Hays, James and Perona, Pietro and Ramanan, Deva and Doll{\'a}r, Piotr and Zitnick, C. Lawrence},
  title     = {Microsoft {COCO}: Common objects in context},
  booktitle = {Proc. European Conf. Computer Vision (ECCV)},
  pages     = {740--755},
  year      = {2014},
  doi       = {10.1007/978-3-319-10602-1_48}
}

@techreport{ts23501r18,
  author = {{3GPP}},
  title = {System Architecture for the {5G} System ({5GS})},
  institution = {ETSI},
  number = {TS 23.501, version 18.8.0, Release 18},
  year = {2025},
  month = jan,
  note = {Clauses 5.7.7, 5.27.2, and 5.37.5.},
  url = {https://www.etsi.org/deliver/etsi_ts/123500_123599/123501/18.08.00_60/ts_123501v180800p.pdf}
}

@article{elkael2026auguste,
  author  = {Elkael, Maxime and Polese, Michele and Lee, Yunseong and Furueda, Koichiro and Melodia, Tommaso},
  title   = {{AUGUSTE}: Online-Learning {dApp} for Predictive {URLLC} Scheduling},
  journal = {arXiv preprint arXiv:2606.03664},
  year    = {2026},
  doi     = {10.48550/arXiv.2606.03664}
}

@article{lee2026kvhandover,
  author  = {Lee, Seunghun and Park, Jihong and Zheng, Ce and Park, Hyuncheol},
  title   = {Low-Latency Edge {LLM} Handover via Joint {KV} Cache Transfer and Token Prefill},
  journal = {arXiv preprint arXiv:2603.28018},
  year    = {2026},
  doi     = {10.48550/arXiv.2603.28018}
}

@misc{qwen2024config,
  author = {{Qwen Team}},
  title = {{Qwen2.5-7B-Instruct}: Model Configuration},
  howpublished = {Hugging Face model repository},
  year = {2024},
  url = {https://huggingface.co/Qwen/Qwen2.5-7B-Instruct/blob/main/config.json},
  note = {Accessed September 6, 2026}
}

@misc{alriksson2020critical,
  author = {Alriksson, Fredrik and Bostr{\"o}m, Lisa and Sachs, Joachim and Wang, Y.-P. Eric and Zaidi, Ali},
  title = {Critical {IoT} Connectivity: Ideal for Time-Critical Communications},
  howpublished = {Ericsson Technology Review},
  year = {2020},
  url = {https://www.ericsson.com/en/reports-and-papers/ericsson-technology-review/articles/critical-iot-connectivity},
  note = {Accessed September 6, 2026}
}
\end{document}